\documentclass{aa}
\usepackage{graphicx,natbib}
\usepackage{txfonts}
\usepackage{xcolor}

\newcommand{\eg} {{\em e.g.}}

\newcommand{\radec}[6]{\emr{\alpha_{2000}=#1^{h}#2^{m}#3^{s}}, \emr{\delta_{2000}=#4^{\circ}#5^{'}#6^{''}}}

\newcommand{\emm}[1]{\ensuremath{#1}}   
\newcommand{\emr}[1]{\emm{\mathrm{#1}}} 
\newcommand{\unit}[1]{\emm{\, \emr{#1}}}
\newcommand{\HHCO}{\mbox{H$_2$CO}}       
\newcommand{\CHtOH}{\mbox{CH$_3$OH}}     

\newcommand{\K}   {\unit{K}}

\newcommand{\mm}  {\unit{mm}}
\newcommand{\m }  {\unit{m}}

\newcommand{\pscm}{\unit{cm^{-2}}}
\newcommand{\pccm}{\unit{cm^{-3}}}

\newcommand{\kms}   {\unit{km\,s^{-1}}}
\newcommand{\Kkms}{\unit{K\,km\,s^{-1}}}
\newcommand{\MHz} {\unit{MHz}}
\newcommand{\GHz} {\unit{GHz}}
\newcommand{\kHz} {\unit{kHz}}

\renewcommand{\deg}{\emm{^\circ}}

\newcommand{\Tsys}{\emm{T_\emr{sys}}}
\newcommand{\Tas}{\emr{T_{A}^{*}}}
\newcommand{\Tmb}{\emr{T_{mb}}}
\newcommand{\Feff}{\emr{F_{eff}}}
\newcommand{\Beff}{\emr{B_{eff}}}

\newcommand{\TabObservations}{%
  \begin{table*}
    \label{tab:obs}
    \caption{Observation parameters. The projection center
      for  all observations is \radec{03}{33}{21.300}{31}{07}{34.00},
      a position in between B1b-N and B1b-S.}
    \begin{center}
      \small
   \begin{tabular}{lcccccccccc}
        \hline
        \hline
        Molecule & Frequency & Instr. & Config. & Beam & PA     & Vel. res. & Int. Time\tablefootmark{a} & \Tsys{} & Noise\tablefootmark{b} & Obs. date \\
                 & \GHz{}    &        &         & $''$ & $\deg$ & \kms{}    & hr                         & K       & mK &\\
        \hline
        \CHtOH{}\tablefootmark{d} & 145.1 & PdBI & C\&D & $2.30\times2.24$ & 128 & 0.16 & 8.8/29 & 160 & 313 & Aug. \& Oct. 2013\\
        \HHCO{}\tablefootmark{d}  & 145.6 & PdBI & C\&D & $2.34\times2.24$ & 108 & 0.20 & 8.8/29 & 160 & 288 & Aug. \& Oct. 2013 \\
        \hline
      \end{tabular}
     \begin{tabular}{lcccccccccc}
        \hline
        Molecule & Frequency & Instr. & \Feff{} & \Beff{} & Res. & Res.   & Int. Time\tablefootmark{c} & \Tsys{} & Noise & Obs. date \\
                 & \GHz{}    &        &         &         & $''$ & \kms{} & hr                         & K       & mK &\\
        \hline
        \CHtOH{} & 145.1 & E150 & 0.93 & 0.74 & 17.9 & 0.16 & 14 & 99 & 44 & Dec. 3rd \& 4th, 2013\\
        \HHCO{}  & 145.6 & E150 & 0.93 & 0.74 & 17.8 & 0.20 & 14 & 99 & 40 & Dec. 3rd \& 4th, 2013\\
        \hline
      \end{tabular}
      \tablefoot{%
        \tablefoottext{a}{Listed as on-source time/telescope time.}%
        \tablefoottext{b}{Evaluated at the mosaic phase center (the noise
          steeply increases at the mosaic edges after correction for
          primary beam attenuation).} %
        \tablefoottext{c}{Telescope time.}
\tablefoottext{d}{FITS files of the H$_2$CO and CH$_3$OH mosaics
 are available in electronic form at the CDS via anonymous ftp to cdsarc.u-strasbg.fr (130.79.128.5)
or via http://cdsweb.u-strasbg.fr/cgi-bin/qcat?J/A+A/.}}
    \end{center}
  \end{table*}}

\begin{document}


\title{{Nascent bipolar outflows associated with the first hydrostatic core candidates  
Barnard 1b-N and 1b-S \thanks{Based on
      observations carried out with the IRAM Plateau de Bure
      Interferometer. IRAM is supported by INSU/CNRS (France), MPG
      (Germany) and IGN (Spain).}
}}

\author{%
M. Gerin \inst{1,2} %
\and J. Pety \inst{3,1} %
\and A. Fuente \inst{4} %
\and J. Cernicharo \inst{5} %
\and B. Commer\c con \inst{6} %
\and N. Marcelino \inst{7} %
}

\institute{%
  LERMA, Observatoire de Paris, CNRS UMR8112, Ecole Normale
  Sup\'erieure, PSL research university,
  24 Rue Lhomond, 75231 Paris cedex 05, France.
  \email{maryvonne.gerin@ens.fr} %
  \and Sorbonne Universit\'es, UPMC universit\'e Paris 06, Paris, France. %
  \and
  Institut de Radioastronomie Millim\'etrique (IRAM),
  300 rue de la Piscine, 38406 Saint Martin d'H\`eres, France.
  \and
Observatorio Astron\'omico Nacional (OAN,IGN), Apdo 112, E-28803 
Alcal\'a de Henares, Spain.
  \and Instituto de Ciencia de Materiales de Madrid (ICMM-CSIC). E-28049,
  Cantoblanco,  Madrid, Spain.
  \and Centre de Recherche Astronomique de Lyon (CRAL), Ecole Normale
  Sup\'erieure de   Lyon, CNRS-UMR5574, France.
  \and
  INAF, Istituto di Radioastronomia, via P. Gobetti 101, 40129 Bologna,
  Italy.
}

\date{Received ; accepted}

\abstract { %
In the theory of star formation, the first hydrostatic core
  (FHSC) phase is a critical step in which a condensed object emerges from
a prestellar core. This step lasts about one thousand years, a very short time
  compared with the lifetime of prestellar cores, and therefore is
  hard to detect unambiguously. 
We present IRAM Plateau de Bure observations of the Barnard 1b dense
  molecular core, combining detections of \HHCO{} and
  \CHtOH{} spectral lines and dust continuum  at 2.3'' resolution ($\sim 500$~AU). %
The two compact cores B1b-N and B1b-S are detected in the dust continuum
  at 2mm, with fluxes that agree with their spectral energy
  distribution.  Molecular outflows associated with both
  cores are detected. They are inclined relative to the direction of the
  magnetic field, in agreement with predictions of collapse in turbulent
and magnetized gas with a ratio of mass to magnetic flux somewhat 
higher than the critical value, $\mu \sim 2 - 7$.  
The outflow associated with B1b-S presents sharp spatial
 structures, with ejection velocities of up to $\sim 7$ \kms\
   from the mean velocity. Its dynamical age is estimated to be $\sim
  2000$~yr. The B1b-N outflow is smaller and slower, with a short dynamical age   of $\sim 1000$~yr.  
The B1b-N outflow mass, mass-loss rate, and mechanical luminosity agree well 
with theoretical predictions of FHSC. 
  These observations confirm the early evolutionary
  stage of B1b-N and the slightly more evolved stage of B1b-S. 
}

\keywords{ISM : clouds; ISM : jets and outflows; ISM : individual objects :
  Barnard 1b; stars : formation}

\maketitle

\section{Introduction}

In the current theory of low-mass star formation, the first condensed
object to be formed in the collapse of the prestellar cores is the first
hydrostatic core (FHSC). During this rather short-lived phase of less than
about a thousand years, the collapse slows down until the core temperature
rises above about 2000\K{}, when molecular hydrogen starts to
dissociate \citep{larson}.  Predictions of the spectral energy distribution
of FHSCs as well as  their density structure and kinematics are available
thanks to  magnetohydrodynamic (MHD) simulations \citep[e.g.][]{commercon:12a,commercon:12b}.  
Their observational
characteristics are intermediate between those of a prestellar core and a
class 0 protostar: i) a spectral energy distribution (SED) peaking beyond
 $\sim 100~\mu$m similar to that of a prestellar core, and ii) a compact
structure of $\sim 500$~AU. 
These features are found in very low luminosity objects (VeLLOs),
which are compact sources characterized by an internal luminosity lower than 0.1 L$_\sun$ 
\citep{difrancesco:07}. They can be class 0 or class I young
stellar objects (YSOs), proto-brown dwarfs, or FHSC candidates. 
VeLLOs exhibit contrasting outflow properties, from extended and collimated lobes down to faint, 
slow, and compact flows (\eg{} \citet{dunham:11}).  The
characterization of an FHSC candidate requires studying its SED and the gas kinematics 
of its envelope and  molecular outflow.
Another important theoretical prediction of the FHSC stage is the presence of
a compact, slow, and poorly collimated outflow, whose properties depend on the strength
and orientation of the magnetic field relative to the rotation axis of the
prestellar core \citep{hennebelle:08,tomida:10,ciardi:10,machida:14}.

As of today, several FHSC candidates have been found in star-forming regions in the  Perseus   
molecular cloud. Of these, the Barnard 1 region 
has attracted attention because of it has dense cores at different evolutionary stages:
B1a and B1c
   are each hosting a class 0 YSO with developed outflows
  \citep{hatchell:07}, while  B1b has recently been proposed to
  host two FHSCs based on the SED analysis of {\sl Herschel} and {\sl Spitzer} data \citep{pezzuto:12}.
 In detail, three remarkable sources are detected in B1b: an infrared
   source detected by {\sl Spitzer}, with strong absorption from ices 
B1b-W   \citep{jorgensen:06,evans:09}, 
and two compact (sub)millimeter continuum sources B1b-N and
  B1b-S that are the FHSC candidates.  The three sources are deeply
  embedded in the surrounding protostellar core, which seems essentially
  unaffected by them, as shown by its high column density,
  $N(\rm{H_2}) \sim 10^{23}$~cm$^{-2}$ and low kinetic temperature, 
$T_K = 12$~\K \ \citep{lis:10}. 

Because of the nearby B1a and B1c YSOs, it has been difficult to associate molecular outflows with
  either B1b-N or B1b-S. \citet{hatchell:07} found high-velocity
  emission close to B1b-S in their CO(3-2) survey of the region, but without a
  clear association with either source. Broad line profiles and line wings
   have also been detected in some species such as CH$_3$OH \citep{oberg:10}
  and the excited transition NH$_3$(3,3) \citep{lis:10}, without assignment
  to any of the B1b sources.
New SMA observation of CO(2-1)
 at a resolution of $6.2'' \times 3.9''$ 
  unambiguously detected molecular outflows with the 
   B1b-S and B1b-W and possibly B1b-N  \citep{hirano:14}.

To better understand the nature of the three objects embedded in
  B1b, we have obtained $2''$ resolution data with the IRAM 
Plateau de Bure
  interferometer, targeting  methanol and formaldehyde lines in the 2 mm
  spectral window. Following  \citet{hirano:14}, we adopt a
  distance of 230 pc for Barnard 1, implying that $2''$ corresponds to
  $\sim 500$\,AU. The observations are
presented in Sect. \ref{sect:obs} and are discussed in Sect.
\ref{sect:results} with particular emphasis on the molecular outflows.

\begin{figure*}
  \rotatebox{0}{ \resizebox{11cm}{!}{
       \includegraphics{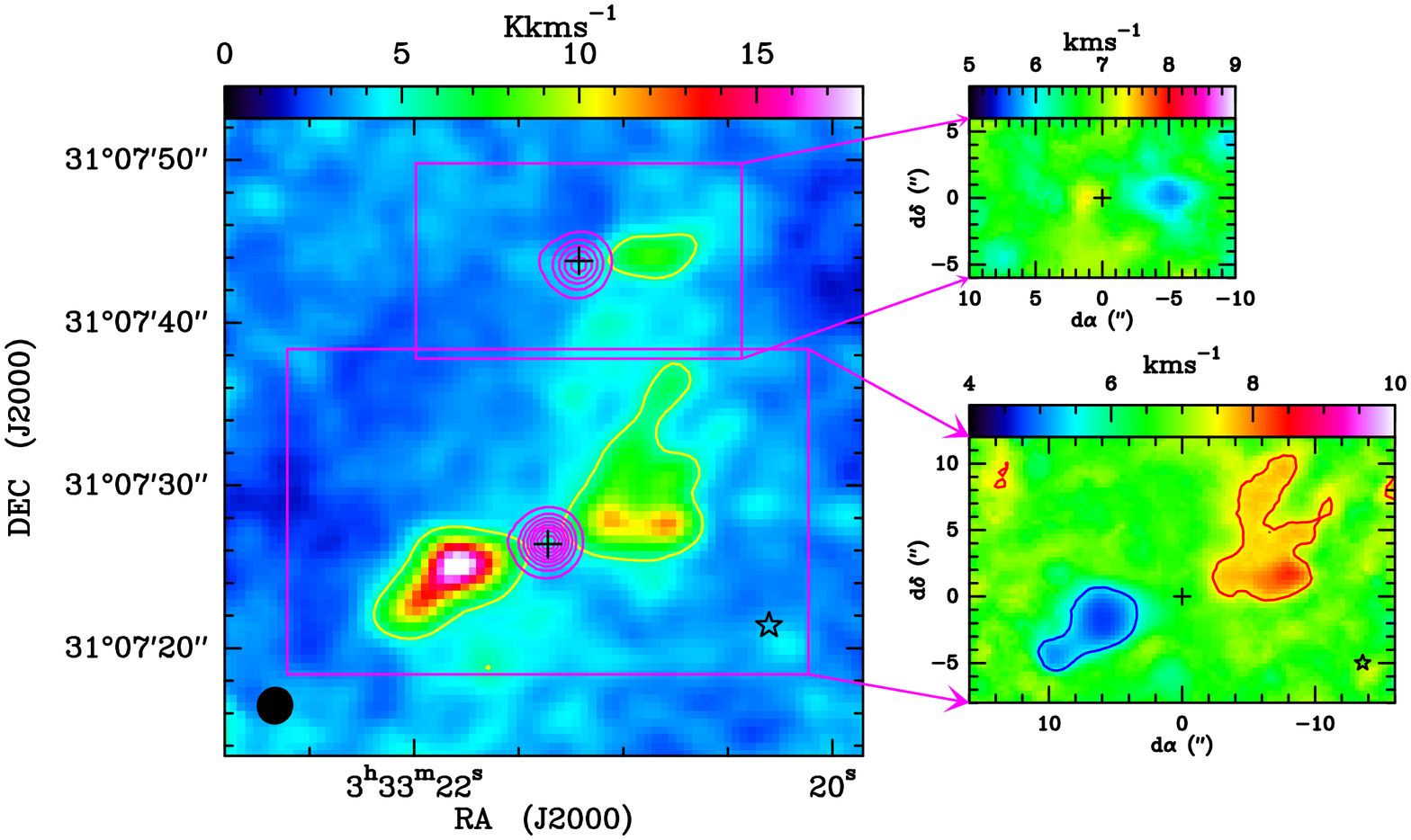}}} %
  \rotatebox{0}{ \resizebox{7cm}{!}{ 
      \includegraphics{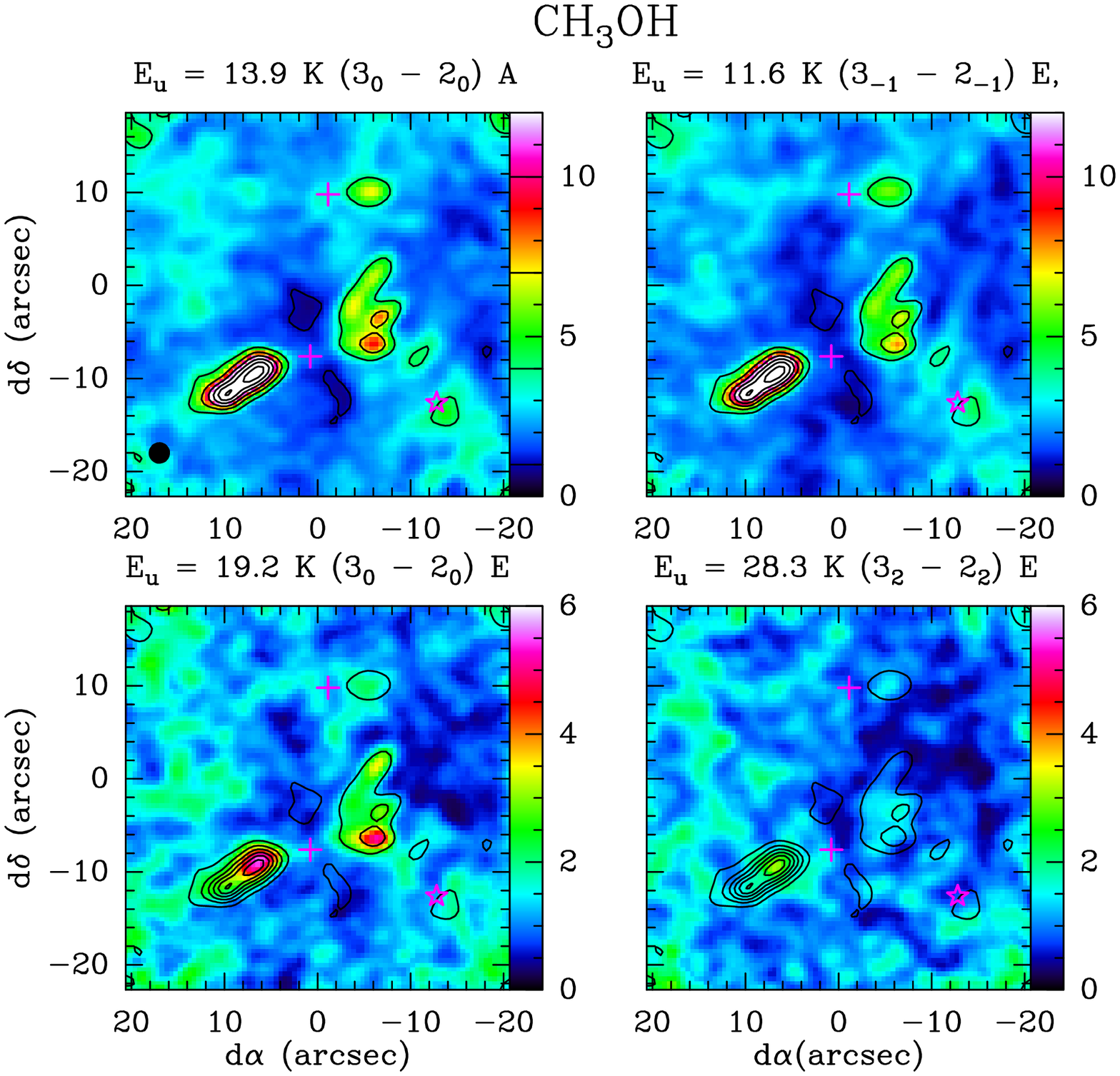}}}
\caption{\label{fig:h2co} Left: Integrated intensity (from 0 to 14 \kms) and
    velocity field of the \HHCO($2_{0,2} - 1_{0,1}$) line.  
Purple contours show the continuum emission at 145~GHz. The plus signs indicate the
    positions of B1b-N and B1b-S. The star shows the position of B1b-W. The black ellipse
shows the beam size.    The small panels show the centroid velocity towards B1b-N (top) and B1b-S
    (bottom) in \kms. Right: 
 Integrated intensity 
images (from 1.5 to 12.5 \kms) of the four
    \CHtOH{} lines detected towards B1b: from left to right and top
    to bottom: $3_{0}-2_{0}$ A, $3_{-1} - 2_{-1}$ E, $3_{0} - 2_{0}$ E and
    $3_{2} - 2_{2}$ E.    
The  intensity scale in \Kkms \ is given
    by the color bar on the right. 
    Contours of the $3_{0}-2_{0}$ A line are superimposed
    in all plots, ranging from 1 to 20\Kkms{} with a step of 3\Kkms. 
Offsets are given relative to the pointing center given in Table 1.
}
\end{figure*}

\section{Observations}
\label{sect:obs}

\TabObservations{} %

Table 1 summarizes the interferometric and single-dish
observations. The calibration and the joint imaging and
deconvolution processes are described in Appendix~\ref{sec:processing}.

\subsection{Interferometric observations}

For the interferometric observations, the lower sideband of the 2\mm{}
receivers was tuned at 145.3\GHz{}. The \texttt{WIDEX} backend yielded a
total bandwith of 4\GHz{} per polarization at a spectral channel spacing of
2\MHz. The intermediate frequency was further split into two 1\GHz{} chunks
centered around 144.5 and 145.3\GHz{}. The eight windows  (40\MHz{}) of the
high spectral resolution  correlator were then
centered around potential lines inside the previously defined
\texttt{WIDEX} chunks. This yielded spectra with a 78\kHz{} channel spacing
that we further smoothed to reach a spectral resolution of 0.2\kms{}
(or typically 97\kHz). We targeted   \HHCO ($2_{0,2} - 1_{0,1}$) at 145.60295~GHz  and the band of \CHtOH ($3 - 2$) lines with frequencies ranging 
from 145.0937~GHz to 145.1318~GHz.
We observed a mosaic of seven pointings that followed a hexagonal compact
pattern with nearest neighbors separated by half the primary beam, providing
 a roughly circular field of view of $70''$ diameter at
2\mm.

\subsection{Single-dish observations}

The IRAM-30m data were  simultaneously observed
at 3 and 2\,mm with a combination of the EMIR receivers and the Fourier
transform spectrometers, which yields a bandwidth of 1.8\GHz{} per sideband
per polarization  at a channel spacing of 49\kHz{}. The mixers were
tuned at 85.55 and 144.90\GHz{}.
We used the on-the-fly scanning strategy with a dump time of 0.5~s
and a scanning speed of $10.9''/$s to ensure a sampling of at least three dumps
per beam at $17.8''$, the resolution at 145~GHz. The
$175''\times175''$ map was covered using successive orthogonal scans along
the RA and DEC axes. The separation between two successive rasters was
$6.5''$ ($\sim \lambda/2D$) to ensure Nyquist sampling at the highest
observed frequency. A common reference position located
at offsets $(200'',150'')$ was observed for 10~s every 77.5~s. 
The typical IRAM-30m position accuracy is $\sim 3''$.

\section{Results}
\label{sect:results}

The median noise level achieved over the mosaic is 0.24\K{} (\Tmb{})
in  0.2\kms{} channels and for a typical resolution of $2.3''$.

\begin{table}
  \caption{\label{tab:points}Physical parameters derived from the CH$_3$OH emission
}
  \centering %
  \small{ %
    \begin{tabular}{lccccc}
      \hline
      \hline
      Position\tablefootmark{a} & $\int T dv$\tablefootmark{b}  & $N_E$\tablefootmark{c} & $n(\rm{H}_2)$ & $T_K$ & [CH$_3$OH]
 \\
      & \Kkms & $10^{13}$\pscm & 10$^5$\pccm & \K & 10$^{-8}$\\
      \hline
      B1b-S & \\
      (6,-2)  & 17.9 & 48   & 6 & 30 &  20 \\
      (7,-2)  & 12.4 & 50   & 9 & 70 &  13 \\
      (-9,6)  & 1.9  &  6   & 6 & 30 &  2.5 \\
      (-8,5)  & 5.9  & 13   & 4 & 50 &  8 \\
      (-8,10) & 2.9  & 7.2  & 3 & 30 &  6 \\
      (-7,1)  & 4.9  & 14   & 6 & 30 &  5 \\
      \hline
      B1b-N &   \\ 
      (-5,1)  & 0.9  & 2    & 6 & 20 &  0.8\\
      \hline
    \end{tabular}}
  \tablefoot{%
    \tablefoottext{a}{Offsets are given in arcsec relative to each young stellar object.}
    \tablefoottext{b}{Integrated intensity of the E-CH$_3$OH$(3_{-1}-2_{-1})$ line.}
    \tablefoottext{c}{$N_E = N(\mbox{E-\CHtOH})$; $N(\mbox{\CHtOH}) =  N(\mbox{E-\CHtOH})+ N(\mbox{A-\CHtOH}) \sim 2 \times N_E$.}
}
\end{table}

\begin{table}
  \caption{\label{tab:outflow}Physical properties of the molecular
    outflows 
}
  \centering %
  \tiny{%
    \begin{tabular}{llcccc}
      \hline
      \hline
      Parameter & Unit & \multicolumn{2}{c}{B1b-N} & \multicolumn{2}{c}{B1b-S} \\
      & & Blue\tablefootmark{a} & Red\tablefootmark{b} & Blue\tablefootmark{c} & Red\tablefootmark{d} \\
      \hline
      Max. velocity    & \kms{}                      & 4.5 & 4 & 7.6 & 6.6 \\
      Size             & 10$^3$\,AU                  & 1.3 & 0.6 & 3 & 3.3  \\
      Dyn. time        & 10$^3$\,yr                  & 2.0 & 1.0 & 1.9 & 2.5 \\
      Mass             & 10$^{-3}$\,M$_\sun$           & 1.2 & 0.4 & 3.0 & 1.9 \\
      Momemtum         & 10$^{-3}$\,M$_\sun$\kms{}     & 3.6 & 1.3 & 13 & 6.4 \\ 
      Mass-loss rate   & 10$^{-7}$\,M$_\sun$\,yr$^{-1}$ & 5.3 & 3.8 & 16 & 7.5\\ 
      Mom. flux        & 10$^{-6}$\,M$_\sun$\kms\,yr$^{-1}$ & 1.5 & 1.0 & 6.7 & 2.5\\ 
      Mech. luminosity & 10$^{-3}$\,L$_\sun$           & 0.8 & 0.5 & 8.1 & 3.1\\
      \hline
    \end{tabular}}
  \tablefoot{%
    \tablefoottext{a}{0.5 -- 5.5 \kms,}
 \tablefoottext{b}{7.5 -- 12.5 \kms,}
 \tablefoottext{c}{-4 -- 5.5 \kms,}
 \tablefoottext{d}{7.5 -- 15 \kms.}
}
\end{table}

\begin{figure}
\centering
   \rotatebox{0}{ \resizebox{8cm}{!}{
       \includegraphics{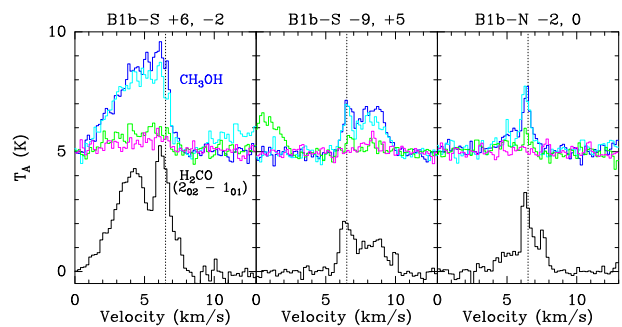}}}
  \caption{\label{fig:spec}
 Examples of \CHtOH{} and
      \HHCO{} spectra towards the B1b-S and B1b-N  outflows. The \HHCO($2_{02}-1_{01}$) line is shown in black. 
    The     four detected \CHtOH{} lines are shown in color: blue $3_0-2_0$ A, cyan
    $3_{-1}-2_{-1}$ E, green $3_0-2_0$ E, and purple
    $3_2-2_2$ E. They have been vertically shifted by 5\K{}
    for clarity.
  }
\end{figure}

\subsection{Geometrical properties of the outflows}
Figure~\ref{fig:h2co} shows the \HHCO ($2_{0,2} - 1_{0,1}$) and \CHtOH{} integrated
intensity maps and the mean velocity field around B1b-N and B1b-S.
The line profiles at selected positions are shown in Fig. \ref{fig:spec}. 
B1b-W is associated with a secondary peak of the \HHCO{} and \CHtOH{}
emission, with narrow line profiles.  
We did not detect any high-velocity emission that might be associated 
with B1b-W or  with the high-velocity CO knot detected by \citet{hirano:14}. The region between B1b-N and B1b-S
presents lower \CHtOH{} emission despite the high dust column density \citep{daniel:13}, a 
behavior characteristic of depletions due to freezing of molecules onto grains, and supported by the detection of 
high column densities of ices 
including \CHtOH{} towards B1b-W \citep{boogert}.

The main result is the clear detection of molecular outflows  in both 
\HHCO{} and \CHtOH{} around B1b-N and B1b-S. 
The red and blue lobes of the B1b-N outflow 
are marginally resolved, but the excellent correspondence
of the \HHCO{} and \CHtOH{} line profiles (Fig. \ref{fig:spec}), and the alignment 
with the B1b-N position support the assignment. 
 The outflow of B1b-S is more extended, with its red and blue lobes
  detected over $\sim 10''$. They are well separated,
 indicating that the inclination angle along the line of sight, $i$,  
is located between $\alpha$ and
$90^\circ - \alpha$,  where $\alpha$ is the opening angle
of either lobe. Using the more extended red lobe, we determine
$\alpha = 23^\circ \pm 8^\circ$, leading to $17^\circ \leq i \leq 73^\circ$.
The red and blue lobes of
the molecular outflows associated with B1b-N and B1b-S appear on 
different sides of these sources, indicating that the outflows 
have different spatial orientations. 

 The
line-of-sight component of the magnetic field towards Barnard~1 was
determined to be $B_{//} = -27 \pm 4 \ \mu$G from OH Zeeman observations 
in a 2.9' beam, toward a position close to IRAS~03301+3057 
and $\sim 43''$ from B1b  \citep{goodman}.
The projected direction of the magnetic field is known from polarimetric
observations of the 850 $\mu$m dust thermal emission with the JCMT at
$\sim15''$ resolution \citep{matthews:02, matthews:09}. It presents
 a regular pattern with a mean N-S
orientation at $PA = 1^\circ \pm 19^\circ$.  
Hence we conclude that
the magnetic field lies in the vertical plane containing the line of sight
and the declination axis, its exact orientation depending on the relative
strengths of its parallel and perpendicular components. Therefore, the
outflows associated with B1b-N and B1b-S are not aligned with the magnetic
field, but present a significant angle with its mean direction.  Using the
available constraints on the orientation (position angle, and for B1b-S, inclination),
and assuming that the component of the magnetic field in the plane of the
sky lies between $19 \ \mu$G as derived by \citet{matthews:02} and 
$100 \ \mu$G, we
constrained the angle $\beta $
 between each outflow and the magnetic field 
 \footnote{The outflow orientation is defined by its projected axis
 position angle $PA$ and  inclination $i$. $\beta $  can be obtained using
  $cos(\beta) = \frac{B_{\perp} cos(PA) cos(i) + B_{//} sin(i)}{\sqrt{B_{\perp}^2+B_{//}^2}} $.} 
to be $\beta > 36^\circ$ for B1b-N and $\beta > 29^\circ$ for B1b-S.
As shown in MHD simulations \citep{ciardi:10}, significant
asymmetries are predicted between the red and blue lobes at such high
inclinations, which may explain the shape of  B1b-N and B1b-S outflows.

\subsection{Physical properties of the outflows}

 We extracted spectra of the four detected \CHtOH{} lines at selected positions given in Table~\ref{tab:points} and
fitted these profiles with a two-component Gaussian, assuming the same
velocity profile for all \CHtOH{} lines. The narrow component centered near 
6.5 \kms \ traces the
quiescent component, while the broad component describes the molecular
outflow.  The line intensities were compared with predictions from 
the non-LTE radiative transfer code \texttt{MADEX} \citep{cernicharo:m} using  collision cross-sections 
from \citet{rabli:10} to derive $n(\rm{H}_2)$, $T_K$ and
\CHtOH{} column densities $N$ in the broad component.
The uncertainties are $ 5 - 10$~\K \  for $T_K$, $\sim 50 \%$ for $n(\rm{H}_2$), 
and $\sim 20\%$ for $N$, as shown by the scatter at nearby positions.
Because the outflows show structures down to the beam size in
the integrated intensity maps (Fig.\ref{fig:h2co}),
we assumed that the length along the line of sight for the 
high-velocity gas is equal to the beam size, $2.3''$ or 530\,AU, to derive
a lower limit to the H$_2$ column density.  This assumption led to
\CHtOH{} abundances relative to H$_2$ in the $10^{-8} - 10^{-7}$ range,
which are intermediate between the high values, $\sim 10^{-5}$, reached in
chemically-rich outflows  \citep{bachiller-perez} 
and the low values measured in quiescent
gas, $\sim 10^{-10}$ \citep{guzman:13}.
The H$_2$ densities derived from MADEX  range from $3 \times 10^5$ to $9 \times 10^5$ \pccm. 
They are comparable with the densities in the B1b envelope determined from the dust continuum emission by \citet{daniel:13}.  
Although higher than the kinetic temperature of the quiescent gas, 12\K{}
\citep{lis:10}, the kinetic temperatures in the shocked gas remain
moderate, between 20 and 70\K.

Using the average properties adapted to each outflow, [\CHtOH] = $7 \times
10^{-8}$ in B1b-S, and  $10^{-8}$ for B1b-N, we computed
the mass in each lobe of the outflowing gas by integrating the emission in
the velocity intervals listed in Table \ref{tab:outflow}. The resulting values are
displayed in Table~\ref{tab:outflow}, together with other parameters.
The outflow masses agree  within a factor of two 
with the previous determinations
based on CO(2-1) \citep{hirano:14}, except for the blue
lobe of the B1b-S outflow, where the CO emission is optically thick.
   The better sensitivity and angular
resolution allow us to obtain lower values for the dynamical time and outflow
size, resulting in higher figures for the momentum and momentum flux by a
factor of $\approx 5$. 
The mechanical luminosities are at least a factor of 
ten lower than the bolometric luminosity of each source.

\subsection{Evolutionary stage of B1b-N and B1b}
B1b-N and B1b-S are located to the right of the class 0 YSOs in the diagram of 
normalized momentum flux, $Fc/L$ versus normalized envelope mass 
$M_{env}/L^{0.6}$  \citep{bontemps:96}, 
with higher normalized envelope masses ($>1.1$  and $>0.7$) 
and a fairly similar normalized 
momentum flux (800 and 1600). Their  luminosities are significantly lower than predicted by the scaling with either the
 envelope mass or the momentum flux, however; this is a typical feature of VeLLOs. 
B1b-N and B1b-S lie in fairly massive cores with at least 0.36 M$_\sun$ in
250 AU \citep{hirano:14}. With such a mass reservoir, the 
final object is most likely a low-mass star.

Previous theoretical studies have shown that the degree of turbulence, 
characterized 
by the ratio of kinetic to gravitational energy $\epsilon,$ and the strength of the magnetic 
field, characterized by the ratio of the mass to flux ratio, to the critical mass to flux ratio, $\mu$ 
\footnote{$\mu = \frac{M/\phi}{(M/\phi)_c}$ with $(M/\phi)_c = 1/\sqrt{2 \pi G}$}, are two key  
parameters for the evolution of collapsing cores \citep[e.g.][]{joos:12,commercon:12a}. Using the
density profile and turbulent line width determined by \citet{daniel:13} in B1b, we derive $\epsilon \sim 0.3$ and $\mu = 2 - 7$ depending on the strength of the
magnetic field. 
The physical parameters of the B1b-N outflow listed in Table \ref{tab:outflow} 
 agree very well with those of
the low outflow component in numerical models of the early evolution of
protostars \citep{joos:12,joos:13,machida:14}, especially for highly 
magnetized cores.
The significant degree of turbulence could explain the misalignment 
with the magnetic field. As explained by \cite{joos:13}, the localized 
rotational motions induced by turbulence lead to a preferred orientation for
the collapsing core that is not related with the local direction of the magnetic field. 
It could also explain the different orientation of the B1b-N and B1b-S outflows.
Models discussed by \citet{commercon:12a}  predicted the mass within 250~AU to be M$_{250AU}$ = 0.16 M$_\sun$ for a total core
 mass of about 1 M$_\sun$ within 4000~AU.  These figures are 
 compatible with the total mass of  B1b, $\sim  2.5$~M$_\sun$ \citep{daniel:13} 
and the mass of the compact mm sources, 0.36~M$_\sun$ each.
The outflow masses, velocities, and mechanical luminosities predicted by the MHD models 
depend on $\mu$, with $M_{out} = 3.3 \times 10^{-3}$~M$_\sun$, $V_{max} = 4.7$~\kms{}, 
$L_{mech} = 4.1 \times 10^{-3}$~L$_\sun$ for $\mu = 2$ .  
For $\mu=10$, the outflow  mass increases to $M_{out} = 2. \times 10^{-2}$~M$_\sun$, 
the luminosity to  $L_{mech} = 7 \times 10^{-3}$~L$_\sun$, while the maximum
velocity decreases to  $V_{max} = 1.7$~\kms.

 These figures are computed about 1000 years after the formation of the FHSC. 
The agreement of the $\mu=2$ case with the properties derived for B1b-N, 
including its short dynamical age, compact size, and cold SED, is excellent, 
supporting its early evolutionary stage.
As already discussed by \citet{hirano:14}, the more energetic outflow properties, higher luminosity, 
and warmer SED of B1b-S only poorly fit with the same models,  providing further
support for the classification of B1b-S as a class 0 YSO, currently in a low accretion stage.

All data so far are consistent with the classification
of B1b-N as an FHSC, but do not provide definitive evidence. 
Higher angular resolution observations are necessary to probe the gas 
dynamics at the $\lesssim 100$AU scale, and a more complete characterization of the chemical and physical properties of B1b is required as well.

\begin{acknowledgements}
  We thank A. Ciardi, P. Hennebelle and D. Lis for illuminating
  discussions, and the referee for a careful review of the manuscript. 
This work was supported by the CNRS program ``Physique
    et Chimie du Milieu Interstellaire'' (PCMI).
We thank the Spanish MINECO for funding support
through grants  AYA2009-07304, AYA2012-32032, FIS2012-32096 and the CONSOLIDER
program "ASTROMOL" CSD2009-00038. The research leading to these results 
has received funding from the European Research Council
under the European Union's Seventh Framework Programme (FP/2007-2013) / ERC-2013-SyG, Grant Agreement n. 610256 NANOCOSMOS. BC aknowledges support by French ANR Retour Postdoc program.
\end{acknowledgements}

\bibliographystyle{aa} %
\bibliography{paper}

\appendix

\section{Data processing}
\label{sec:processing}

\subsection{Interferometric data}

We used the standard algorithms implemented in the software \texttt{GILDAS/CLIC}
 to calibrate the PdBI data. The radio-frequency bandpass was
calibrated by observing the bright (12.8 Jy) quasar 3C84. Phase and
amplitude temporal variations where calibrated by fitting spline
polynomials through regular measurements of two nearby ($<11\degr$) quasars
(3C84 and 0333+321). The PdBI secondary flux calibrator MWC\,349 was
observed once during every track, which allowed us to derive the flux scale
of the interferometric data. The absolute flux accuracy is $\sim 10\%$.

To produce the continuum maps, we imaged and deconvolved the \texttt{WIDEX}
data at 2\MHz{} resolution. This allowed us to identify all the detected
lines and to remove them before synthesizing the continuum image. To subtract the continuum from the lines, we first synthesized continuum uv
tables in a $\sim 400\MHz$ frequency range devoid of lines close
to the
targeted line. This way, we did not need to take a potential
variation of the continuum level with frequency into account.

\subsection{Single-dish data}

Data reduction was carried out using the software \texttt{GILDAS/CLASS}\footnote{See
    \texttt{http://www.iram.fr/IRAMFR/GILDAS} for more information about
    the GILDAS softwares~\citep{pety05}.}
. The data were first calibrated to the \Tas{} scale using the
chopper-wheel method~\citep{penzias73}. The spectra were converted to main-beam
temperatures (\Tmb{}) using the forward and main-beam efficiencies (\Feff{}
and \Beff{}) listed in Table~1. The resulting amplitude
accuracy is 10\%. A 20\MHz{}-wide subset of the spectra was first extracted
around each line rest frequency. We computed the experimental noise after
subtracting a first-order baseline from every spectrum, excluding the
velocity range from 4 to 9\kms{} LSR where the signal resides. A systematic
comparison of this noise value with the theoretical noise computed from the
system temperature, the integration time, and the channel width allowed us
to filter out outlier spectra (typically 3\% of the data). The spectra were
then gridded into a data cube through a convolution with a Gaussian kernel
of FWHM$\sim1/3$ of the IRAM-30m telescope beamwidth.

\subsection{Joint imaging and deconvolution of the interferometric and
  single-dish data}

Following~\citet{rodriguez08}, the  software \texttt{GILDAS/MAPPING}  and the
single-dish map from the IRAM-30m were used to create the short-spacing
visibilities not sampled by the Plateau de Bure interferometer. In short,
the maps were deconvolved from the IRAM-30m beam in the Fourier plane
before multiplication by the PdBI primary beam in the image plane. After a
last Fourier transform, pseudo-visibilities were sampled between 0 and
15\m{}, which is the difference between the diameters of the IRAM-30m and the PdBI
antennas.

These visibilities were then merged with the interferometric
observations. Each mosaic field was imaged and a dirty mosaic was built
by combining these fields in the following optimal way in terms of
signal--to--noise ratio~\citep{pety10}. The dirty intensity distribution was
corrected for primary beam attenuation, which induces a spatially
inhomogeneous noise level. In particular, noise strongly increases near the
edges of the field of view. To limit this effect, both the primary beams
and the resulting dirty mosaics were truncated. The standard level of
truncation was set at 20\% of the maximum in \texttt{GILDAS/MAPPING}. The
dirty image was deconvolved using the standard H\"ogbom CLEAN
algorithm. The resulting data cube was then scaled from Jy/beam to the \Tmb{}
temperature scale using the synthesized beam size (see
Table~1).

The \HHCO{} emission covers most of the mosaic field of view. The
emission structure thus seems to sit on a constant brightness that only
depends on the frequency, not on the spatial position. CLEAN deconvolution
methods have many difficulties to properly deconvolve this ``constant''
emission. To avoid this, we subtracted the mean spectrum over this field of
view of the single-dish data before processing them to produce the
short-spacings. We then imaged and deconvolved the hybrid data set as
explained above and added this mean spectrum back to the hybrid
synthesis data cube (30m + PdBI) after deconvolution and conversion to the
\Tmb{} temperature scale. This treatment is correct because a constant
emission is always resolved, that is, independent of the resolving power of
the observatory.

\section{Spectral energy distribution}

\begin{figure}
   \resizebox{7cm}{!}{
     \includegraphics{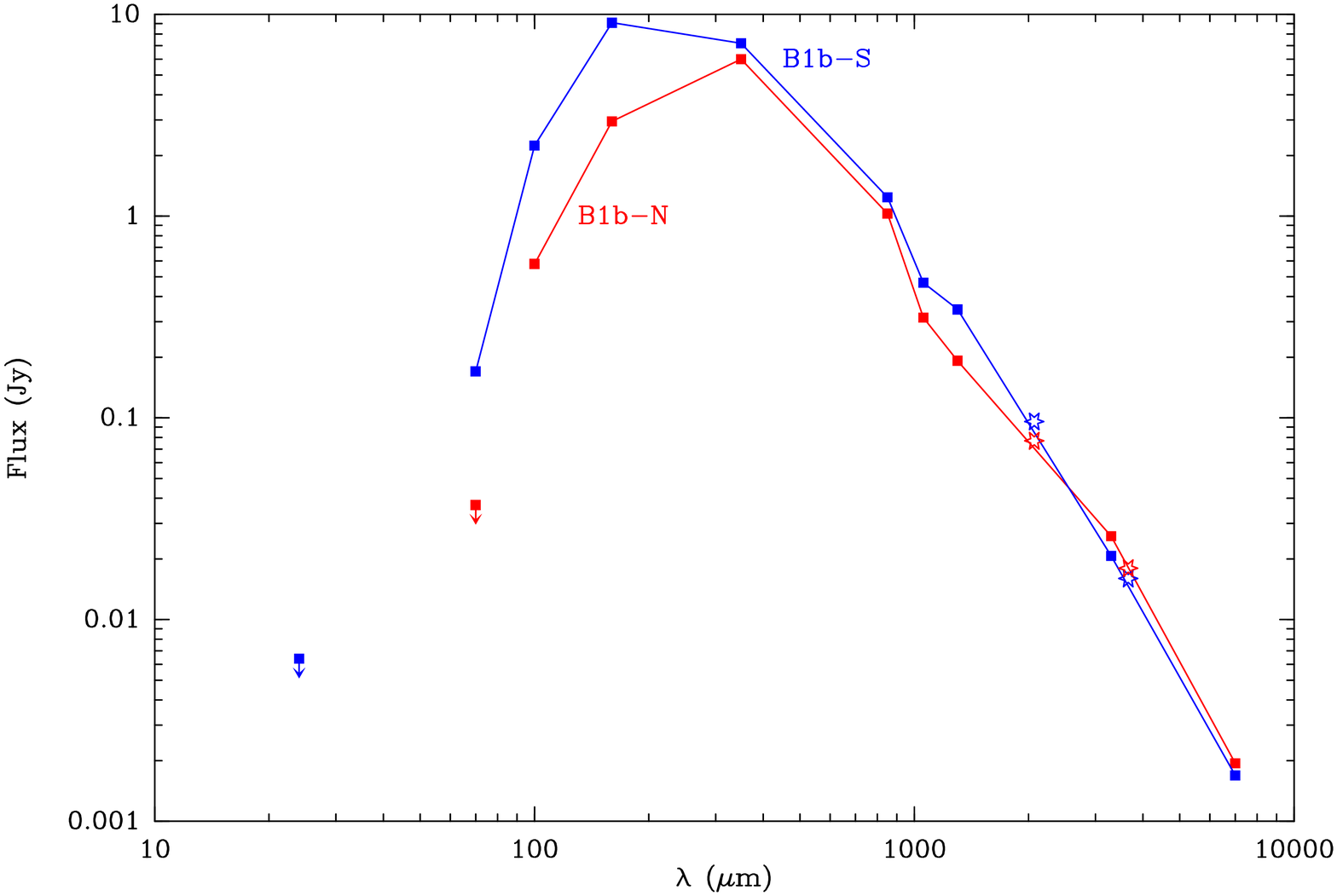}}
  \caption{\label{fig:sed} Spectral energy distribution of B1b-S (blue) and
    B1b-N (red).  The data points are taken from \citet{hirano:14} and
    include {\it Spitzer}, {\it Herschel,} and ground-based observations.
  The stars show the new PdBI
    measurements.  }
\end{figure}

We present in Fig.~\ref{fig:sed} the spectral energy distribution of
  B1b-N and B1b-S and provide refined positions in Table~\ref{tab:sources}.
  The measured continuum fluxes agree well with the values reported by
  \citet{pezzuto:12} and~\citet{hirano:14}. It is interesting to observe
  that B1b-S is more luminous than B1b-N at far-infrared and submillimeter
  wavelengths while the reverse is true long-ward of $\sim 2.5$~mm. The new
  PdBI data help to locate the crossing point of the spectral energy
  distributions.

\begin{table}
  \caption{\label{tab:sources} Positions and fluxes of the protostellar sources
}
  \centering %
  \tiny{ %
  \begin{tabular}{lcccc}
    \hline
    \hline
    Name & RA & Dec & L  & F$_{145}$ \\
    &     &    & (L$\sun$)& (mJy)   \\
    \hline
    B1b-N & 03:33:21.21 & 31:07:43.8 & 0.15 &$77 \pm 4$   \\
    B1b-S & 03:33:21.36 & 31:07:26.4 & 0.31 & $96 \pm 5$    \\
    B1b-W  & 03:33:20.30 & 31:07:21.4 & 0.17/0.25 & $< 4$     \\
    \hline
  \end{tabular}}
\end{table}

\end{document}